\documentstyle[12pt]{article}
\begin{document}
%
%


\textheight 9in
\topmargin -.25in
\oddsidemargin 0in
\evensidemargin 0in

 \jot=4mm
\textwidth=16truecm
\baselineskip=20pt    
\parindent=20pt
\footskip=48pt
\hoffset=-0.9cm  
\oddsidemargin=1cm 
\hsize=15.5truecm  
\setlength{\unitlength}{.1cm}
\pagenumbering{arabic}
\renewcommand{\theequation}{\thesection.\arabic{equation}}


\def\beqa{\begin{eqnarray}}
\def\eeqa{\end{eqnarray}}
\def\beq{\begin{equation}}
\def\eeq{\end{equation}}
\def\beqal{\begin{eqnarray}\label}
\def\beql{\begin{equation}\label}


\def\vol{\int d^4x\,\sqrt{-g}} 
\def\tvol{\int d^4x\,\sqrt{-\tilde{g}}} 
\def\grav{-\frac{1}{16 \pi }}
\def\half{\frac{1}{2}}
\def\gu{g^{\m\n}}
\def\gd{g_{\m\n}}
\def\tgd{\tilde{g}_{\m\n}}
\def\tgu{\tilde{g}^{\m\n}}
\def\lc{\raisebox{-.7ex}{$\stackrel{\textstyle <}{\sim}$}}
\def\gc{\raisebox{-.7ex}{$\stackrel{\textstyle >}{\sim}$}}
\def\R{\mbox{\rm I\kern-.18em R}}
\def\P{\mbox{\rm I\kern-.18em P}}
\def\Ds{\ {\big / \kern-.70em D}}
\def\uno{{1 \kern-.30em 1}}
\def\ds{\big / \kern-.90em {\ \p} }

\def\cDs{\ {\big / \kern-.70em {\cal D}}}
\def\Q{{\kern .1em {\raise .47ex \hbox{$\scriptscriptstyle |$}}
\kern -.35em {\rm Q}}}
\def\de{\mbox{\it d}}
\def\Tr{\mbox{\rm Tr}}
\def\Im{\mbox{\rm Im}}
\def\cl{\mbox{\scriptstyle cl}}
\def\tr{\mbox{\rm tr}}
\def\ap{\alpha^{\prime}}
\def\psb{\bar{\psi}}
\def\chb{\bar{\chi}}
\def\lb{\bar{\lambda}}
\def\epsb{\bar{\varepsilon}}
\def\sb{\bar{\sigma}}
\def\ad{\dot{\alpha}}
\def\bd{\dot{\beta}}
\def\wb{\bar{w}}
\def\vb{\bar{v}}
\def\ib{\bar\imath}
\def\jb{\bar\jmath}
\def\um{^{\m}}
\def\un{^{\n}}
\def\dm{_{\m}}  
\def\dn{_{\n}}
\def\umn{^{\m\n}}
\def\dmn{_{\m\n}}
\def\umnrs{^{\m\n\r\s}}
\def\dmnrs{_{\m\n\r\s}}
\def\ua{^{\al}}  
\def\ub{^{\b}}
\def\da{_{\al}}
\def\db{_{\b}}
\def\ug{^{\g}}
\def\dg{_{\g}}
\def\uam{^{\al\m}}
\def\uan{^{\al\n}}
\def\uab{^{\al\b}}
\def\dab{_{\al\b}}
\def\dabgd{_{\al\b\g\d}}
\def\uabgd{^{\al\b\g\d}}
\def\udeab{^{;\al\b}}
\def\ddeab{_{;\al\b}}
\def\ddemunu{_{;\m\n}}
\def\udemunu{^{;\m\n}}
\def\ddemu{_{;\m}}  \def\udemu{^{;\m}}
\def\ddenu{_{;\n}}  \def\udenu{^{;\n}}
\def\ddea{_{;\al}}  \def\udea{^{;\al}}
\def\ddeb{_{;\b}}  \def\udeb{^{;\b}}

\def\naba{\nabla_{\a}}
\def\nabb{\nabla_{\b}}
\def\nabm{\nabla_{\m}}
\def\nabn{\nabla_{\n}}
\def\pmu{\partial_{\m}}
\def\pn{\partial_{\n}}
\def\p{\partial}

\def\bib#1{$^{\ref{#1}}$}

\def\al{\alpha} 
\def\b{\beta}
\def\g{\gamma}
\def\d{\delta}
\def\eps{\varepsilon}
\def\z{\zeta}
\def\h{\eta}
\def\th{\theta}
\def\k{\kappa}
\def\l{\lambda}
\def\m{\mu} 
\def\n{\nu} 
\def\x{\xi}
\def\r{\rho}
\def\s{\sigma}
\def\t{\tau}
\def\ph{\phi}
\def\ch{\chi}
\def\ps{\psi}
\def\om{\omega}
\def\G{\Gamma}
\def\D{\Delta}
\def\Th{\Theta}
\def\L{\Lambda}
\def\S{\Sigma}
\def\Ph{\Phi}
\def\Ps{\Psi}
\def\O{\Omega}

\def\cf{{\cal F}}
\def\ca{{\cal A}}
\def\cc{{\cal C}}
\def\cg{{\cal G}}
\def\cd{{\cal D}}
\def\cv{{\cal V}}
\def\cm{{\cal M}}
\def\co{{\cal O}}
\def\da{\dot{A}}
\def\db{\dot{B}}
\def\1{\dot{1}}
\def\2{\dot{2}}
\def\dd{\raisebox{+.3ex}{$\stackrel{\scriptstyle \leftrightarrow}{\partial}$}}

\def\a{\`a\ }\def\o{\`o\ }\def\ii{\`\i\ }
\def\u{\`u\  }\def\e{\`e\ }\def\ke{ch\'e\ }

\font\mybb=msbm10 at 12pt
\def\bb#1{\hbox{\mybb#1}}
\font\mmybb=msbm10 at 8pt
\def\bbb#1{\hbox{\mmybb#1}}
\def\Z {\bb{Z}}
\def\Zpic {\bbb{Z}}
\def\R {\bb{R}}
\def\C {\bb{C}}
\def\H {\bb{H}}

\def\real{{\bb{R}}}
\def\rreal{{\bbb{R}}}
\def\rational{\bb{Q}}
\def\R4{\real^4}
\def\Ker{\mbox{Ker}}

\def\jmp{{\rm J. Math. Phys.}\ }
\def\prd{{\rm Phys. Rev. {\bf D}} }
\def\pre{{\rm Phys. Rev.}\ }
\def\jp{{\rm J. Phys.}\ }
\def\prl{{\rm Phys. Rev. Lett.}\ }
\def\pl{{\rm Phys. Lett. {\bf B}}}
\def\npb{{\rm Nucl. Phys. {\bf B}}}
\def\mpl{{\rm Mod. Phys. Lett.}\ }
\def\annp{{\rm Ann. Phys.}\ }
\def\ijmp{{\rm Int. J. Mod. Phys.}\ }
\def\cmp{{\rm Comm. Math. Phys.}\ }
\def\cqg{{\rm Class. Quant. Grav.}\ }
\def\spj{{\rm Sov. Phys. JETP}\ }
\def\spjl{{\rm Sov. Phys. JETP lett.}\ }
\def\prs{{\rm Proc. R. Soc.}\ }
\def\grg{{\rm Gen. Rel. Grav.}\ }
\def\nat{{\rm Nature}\ }
\def\apj{{\rm Astrophys. J.}\ }
\def\aaa{{\rm Astron. Astrophys.}\ }
\def\ncim{{\rm Nuovo Cim.}\ }
\def\ptp{{\rm Prog. Theor. Phys.}\ }
\def\aip{{\rm Adv. Phys.}\ }
\def\jpamg{{\rm J. Phys. A: Math. Gen.}\ }
\def\mnras{{\rm Mon. Not. R. Astr. Soc.}\ }
\def\prep{{\rm Phys. Rep.}\ }
\def\rmp{{\rm Rev. Mod. Phys.}\ }
\def\ncb{{\rm Il Nuovo Cimento ``B''}}
\def\ssr{{\rm Space Sci. Rev.}\ }
\def\pasp{{\rm Pub. A. S. P.}\ }
\def\araa{{\rm Ann. Rev. Astr. Ap.}\ }
\def\asr{{\rm Adv. Space Res.}\ }
\def\etal{{\rm et al.}\ }
\def\ie{{\it i.e. }}
\def\eg{{\it e.g. }}

\renewcommand{\baselinestretch}{1.3}
\pagenumbering{arabic}
\setcounter{page}{1}
\thispagestyle{empty}
\vskip 0.5cm
\begin{flushright}
ROM2F--97--47 \\
{\tt hep-th/9712218}
\end{flushright}
\centerline{\large \bf INSTANTON CALCULUS AND SUSY GAUGE THEORIES} 
\vskip 0.2cm
\centerline{\large \bf ON ALE MANIFOLDS}
\vspace{2cm}
\centerline{\sc DIEGO BELLISAI}
\vskip 0.1cm
\centerline{\sl Dipartimento di Fisica, 
Universit\`a di Roma  ``Tor Vergata"}  
\centerline{\sl Via della Ricerca Scientifica, \ \ 00133 \ Roma, \ \ ITALY}
\vskip .8cm
\centerline{\sc GABRIELE TRAVAGLINI}
\vskip 0.1cm
\centerline{\sl I.N.F.N. \ -- \  Sezione di Roma II, }
\centerline{\sl Via della Ricerca Scientifica, \ \ 00133 \ Roma, \ \ ITALY}
\vskip 1.5cm
\centerline{\bf ABSTRACT}
\vskip 0.4cm
{We study instanton 
effects along the Coulomb branch of an $N=2$
supersymmetric Yang--Mills theory with gauge group $SU(2)$ on Asymptotically 
Locally Euclidean (ALE) spaces. We focus our attention on an 
Eguchi--Hanson gravitational background and on gauge field configurations 
of lowest Chern class.}
\vskip .1in
\newpage
\setcounter{page}{1}
\setcounter{equation}{0}
\setcounter{section}{0}
\section{Introduction}
Globally supersymmetric Yang--Mills (SYM) theories on four--manifolds 
\cite{WITTEN} 
provide a natural framework in which to study non--perturbative effects. 
The existence of non--renormalization theorems 
\cite{SIEGEL,SEIBNONREN} allows one to exactly 
compute physical quantities like superpotentials in $N=1$ theories 
\cite{INTRSEIB}. Moreover, in 
$N=2$ SYM theories, holomorphy requirements on the prepotential 
\cite{Gates} are the crucial 
ingredients needed to determine the quantum moduli space and the Wilsonian 
effective action \cite{SW}. 

Instantons \cite{BPST,THO} 
are among the most interesting non--perturbative field configurations. 
In particular, they 
proved to be a fundamental tool for checking, from first principles 
and quantitatively, 
the exactness of the solutions proposed by Seiberg and Witten in 
$N=2$ SYM theory and supersymmetric QCD \cite{DKM}. 
Furthermore, in some theories with matter in chiral representations, 
they are known to trigger dynamical SUSY breaking \cite{susy}. 

It is possible to perform instanton calculations in different phases of 
supersymmetric field theories. 
If the scalar fields of the theory have zero vacuum expectation value,  
instantons are exact saddle points of the action functional around which 
to perform semiclassical approximations. 
If the scalar fields have non--zero 
vacuum expectation values, instantons 
are just approximate solutions of the equations of motion. 
On the other hand, 
when the vacuum expectation values are 
much larger than the renormalization group invariant scale of the 
theory, it is possible to perform reliable instanton calculations in a 
weak--coupling regime (``constrained instanton" method 
\cite{THO,AFF}). 

In the following we will focus our attention on  $N=2$ SYM theories with 
gauge group $SU(2)$. 
In this case, when the complex scalar field has a non--zero 
vacuum expectation value, the gauge group is spontaneously broken 
down to $U(1)$. One can then study the dynamics of the low--energy theory 
which is obtained after integrating out the high--frequency modes. 
The first motivation of our computational approach  consists in studying 
the non--perturbative dynamics of $N=2$ 
effective theories on curved backgrounds via instanton calculus. 
To this end, the correct choice is to give non--zero vacuum expectation values 
to the scalar fields. One could thus infer the instanton corrections 
to the $N=2$ holomorphic prepotential, which encodes the low--energy 
dynamics. 

Among all possible four--manifolds a special class is represented by 
manifolds which have a self--dual Riemann tensor and which are known as 
``Asymptotically Locally Euclidean gravitational instantons", since 
they are solutions of Einstein's 
equations with vanishing gravitational action. 
These manifolds have played a key r\^{o}le in the study of Euclidean 
quantum gravity (for a review see \cite{EHREP}). 
Indeed, similarly to gauge instantons, they induce calculable
non--perturbative 
effects which may cause a dynamical breaking of supersymmetry
\footnote{Generally, on curved manifolds SUSY is not globally realized. 
However, as pointed out in \cite{WITTEN}, in $N=2$ SYM theories there exists a 
conserved scalar supercharge which can be interpreted as the BRST generator 
of the topological symmetry of the twisted version of the theory. 
In particular, this ensures that certain correlators of local operators
are position--independent.}. 
Among ALE gravitational instantons, the simplest and the most 
investigated one is the Eguchi--Hanson solution \cite{EH}. 
The formation of fermionic condensates in this background 
has been studied both in a pure supergravity \cite{KONMAGN,HAW} and 
in an  effective string theory context \cite{BFRMOLD}. 
In particular, in \cite{KONMAGN} the gravitino field--strength condensate 
was explicitly computed and found to be finite and position--independent, 
possibly responsible for local supersymmetry breaking. 
Moreover, from a stringy point of view, ALE manifolds represent 
absolute minima of the gravitational part of the action which is obtained 
as a low--energy limit of the heterotic (and type I) string. 
Our second motivation is that in this context one could 
perform interesting string--inspired calculations, 
and explore a possible supersymmetry breaking
(the underlying string theory acting as a regulator for the 
non--renormalizable supergravity theory)
\footnote{A study of non--perturbative effects in global SYM theories in the 
Eguchi--Hanson background has been performed in the absence of 
vacuum expectation values for the scalar fields in \cite{BFRM2}.}. 
Gauge instantons of $c_{2}=1/2$ and 1 are solutions of the string equations 
of motion to lowest order in the $\s$--model coupling constant 
$\al^{\prime}$. 
The case $c_{2}=3/2$, instead, corresponds to the identification 
of the gauge connection with the spin connection \cite{BFRMOLD} 
(``standard embedding") 
and the solution is conjectured not to get perturbative corrections in 
$\al^{\prime}$. 
As a first step towards the case $c_2=3/2$ 
(which presents formidable computational difficulties)
we start our investigation 
by studying correlation functions in the topological sector $c_2=1/2$,
which could provide us with a useful roadmap for further progress along 
that direction.

The plan of the paper is as follows. 
In section 2 we briefly review $N=2$ SYM theories on the Eguchi--Hanson 
manifold, mainly to fix our conventions. 
In section 3 we examine the (gauge) instanton configuration 
of Chern class $c_{2}=1/2$, around 
which we will expand the generating functional of Green's functions. 
On curved backgrounds, one must pay 
particular attention to the treatment of the collective coordinates which 
describe the instanton orientation in color space. We carefully discuss 
this issue and 
show how to correctly perform the integration over the related moduli 
in order to restore all the unbroken symmetries of the model. 
In this section 
we also collect the bosonic and fermionic zero--mode norms, and compute 
the classical Higgs and Yukawa actions. 
Section 4 is devoted to the (semiclassical) evaluation of some 
instanton--dominated correlators of  the microscopic theory. 
In the final section we draw some conclusions and discuss some further 
issue under investigation. 
\section{Description of the model}
\setcounter{equation}{0}
We intend to study instanton--dominated correlators in $N=2$ 
globally supersymmetric theories in the Eguchi--Hanson 
background.

The Eguchi--Hanson metric is given by
\beq\label{eh}
d s^2=g_{\m\n}dx^{\m}dx^{\n}=
\left(\frac{r}{u}\right)^2 dr^2+r^2 (\s_{x}^2+\s_{y}^2)+u^2 \s_{z}^2
\ \ ,
\eeq
where $\s_{x},\s_{y},\s_{z}$ are the 
left--invariant forms on $SU(2)$ and $u=r\sqrt{1-(a/r)^4}$. 
The metric (\ref{eh}) 
has a bolt singularity at $r=a$ which can be removed by changing to the 
radial variable $u$ and by identifying antipodal points. 
Thus the boundary is $S^{3}/\Z_{2}$;  
moreover, the manifold is not invariant under the action of the Poincar\'e 
group,  
but admits an isometry group which is 
$SU(2)_{R}\otimes U(1)_{L}$. In particular, and this will be crucial in the 
following, the manifold is not translationally invariant since antipodal 
points are identified.

In order to study non--perturbative contributions to Green's functions 
we need to know the 
form of the gauge instantons
\footnote{Or, at least, a parametrization which describes the 
full moduli space of the instanton solution (see \cite{DKMULTIMO}).}. 
On flat space, this can be achieved through the  
Atiyah--Drinfeld--Hitchin--Manin (ADHM) construction  
\cite{ADHM,ATI}, which naturally 
provides us with an algorithm which determines the most general 
self--dual instanton connection. 
Its extension to the case of ALE spaces 
was found by Kronheimer and Nakajima \cite{KRONNAK}.
It was then translated in a more physical language in \cite{BFRM,BFRMLONG}, 
where the explicit expressions of the self--dual connection 
on the minimal instanton bundle ${\cal E}$ (with second Chern 
class $c_{2}({\cal E})=1/2$) and of the bosonic and fermionic zero--modes 
were derived and implicit 
formulas for the cases $c_{2}=1$ and 3/2 were given. 

The $N=2$ Super Yang--Mills action is 
$S_{\rm SYM}=\int d^4 x\sqrt{g}\ L_{\rm SYM}$, where
\beqa
L_{\rm SYM}&=& \frac{2}{g^2} {\rm Tr}\left[ 
\frac{1}{4}F^{\m \n} F_{\m \n}+
i \lb_{A} \Ds \l^{A}+ 
(D^{\m} \phi )^{\dagger} (  D_{\m}\phi ) + \right.
\nonumber\\
& &+\left.\frac{1}{2}[\phi , \phi^{\dagger} ]^2 
+\left(\frac{1}{\sqrt{2}}[\phi^{\dagger}, \l_{A}]\l_{B} \eps^{AB}
 + {\rm h.c.}\right) \right] \ \ ,
\eeqa
$A,B=1,2$ are supersymmetry indices,  and $\eps_{1 2}=
-\eps_{2 1}=1$
\footnote{We choose the generators in the fundamental representation 
to be $T^a=\t^a /2$, $\t^a$ being the Pauli matrices.}. 

The condition for the vacuum state of the theory 
to be $N=2$ supersymmetric is that the 
potential $V(\phi,\phi^{\dagger})$ vanishes, that is
\beq\label{commut}
[\phi,\phi^{\dagger}]=0 \ .
\eeq
This means that the solution of (\ref{commut}) 
is a normal operator, which can therefore be 
diagonalized by an $SU(2)$ color rotation $\O$. Then, the 
classical
\footnote{Actually neither perturbative 
nor non--perturbative quantum corrections can lift the vacuum degeneracy 
\cite{SHIF,SEI}.} (supersymmetric) vacuum configuration for the Higgs field 
can be written as 
\beq\label{mizzica2}
\langle \phi^{a} \rangle =\O^{a}{}_{b}(v \d^{b3})\ \ ,
\eeq
where $v\in\C$ and $a,b=1,2,3$.  
One can then 
expand (\ref{mizzica2}) in the 
basis  
\beq\label{mizzica}
\phi_{0}^{a(b)}=v\d^{ab}\ \ .
\eeq
When $v\neq 0$ the gauge symmetry is spontaneously broken to $U(1)$, and we 
are left with an $N=2$ SUSY abelian low--energy theory. 
\section{The instanton configuration and the choice of 
\newline
collective coordinates}
\setcounter{equation}{0}
The gauge instanton solution for $c_{2}({\cal E})=1/2$  is given by
acting with a global color rotation $R$ on the basic instanton configuration 
\cite{BFRM}
\beq
A=A_{\m}d x^{\m}=i\pmatrix{f(r)\s_{z} & g(r)\s_{-}\cr\cr
g(r)\s_{+} & -f(r)\s_{z}}\ \ ,
\eeq
where $\s_{\pm}=\s_{x}\pm i\s_{y}$, and 
\beq
f(r)=\frac{t^2 r^2+a^4}{r^2 (r^2+t^2)}\ \ ,\qquad 
g(r)=\frac{\sqrt{t^4-a^4}}{r^2+t^2}\ \ .
\eeq
The matrix $R$ contains the collective coordinates related to global color 
rotations. 
When $a=0$ this configuration becomes the 't Hooft instanton in the 
so--called ``singular gauge", centered around the origin. 

To compute instanton--dominated Green's functions one has to know the 
explicit form of the bosonic and fermionic zero--modes. 
In the $N=2$ Eguchi--Hanson background there exist four gaugino zero--modes 
which are related to the fact that the instanton solution 
explicitly breaks the superconformal symmetry. 
They can be written as
\beq\label{boh}
\l_{\al(0)}^{aA}=\s^{\m}_{\al\dot{\al}}(D_{\m}\th)^{a}\epsb^{A\dot{\al}}
\ \ ,
\eeq
where $\epsb^{A\dot{\al}}$ are the two covariantly constant spinors on the 
Eguchi--Hanson background and $\theta_{a}$ (with $a=1,2,3$) are the 
bounded solutions of the scalar Laplace equations 
\beq\label{scalar}
[D^2 (A)]_{a}{}^{b}\th_{b}=0\ \ .
\eeq
These equations  can be recast (with a radial Ansatz)
in the form 
\beqa\label{elena}
& &\left[\frac{1}{r^3}\frac{\p}{\p r}(r u^2) \frac{\p}{\p r}-
4\left(\frac{g^2}{r^2}+\frac{f^2}{u^2}\right)\right]\th_{1,2}=0\ \ ,
\nonumber\\
& &\left[\frac{1}{r^3}\frac{\p}{\p r}(r u^2) \frac{\p}{\p r}-
\frac{8g^2}{r^2}\right]\th_{3}=0\ \ .
\eeqa
The solutions to (\ref{elena}) with boundary conditions 
$\lim_{r\to \infty} \th_{a}(r) = 1$
are given by
\beq\label{scalsol}
\th_{1}=\th_{2}=\frac{\sqrt{r^4-a^4}}{r^2+t^2}\ \ , \qquad 
\th_{3}=\frac{t^2 r^2+a^4}{t^2 (r^2+t^2)}\ \ .
\eeq
The most general form of the gaugino zero--modes  
can be re--written in terms of two constant spinors $\eta^{A}_{\al}$ as 
the global color rotation $R$ acting on the configuration (\ref{boh}), 
that is 
\beq
\label{gaugino}
[\l_{\al(0)}^{aA}]_{R}=R^{a}{}_{b}(f^{b}\s^{b}_{\al}{}^{\b}\eta^{A}_{\b})\ \ ,
\eeq
where 
\beqa
f^{1}&=&f^{2}=\frac{2(t^2 r^2+a^4)}{r(r^2+t^2)^2}\ \ ,\nonumber\\
f^{3}&=&\frac{2\sqrt{(t^4-a^4)(r^4-a^4)}}{r(r^2+t^2)^2}\ \ .
\eeqa
The norm of the gaugino zero--modes is 
\beq
||\l||^2=\int d^4 x\sqrt{g}\ (\l_{\al}^{a})^{\ast}
(\sb^{0})^{\dot{\al}\b}\l_{\b}^{a}=\sqrt{2}\pi t \ \ .
\eeq
The four zero--modes of the gauge field are related to the global 
symmetries broken by the instanton background, {\it i.e.} dilatations and 
$SU(2)$ rotations. 
The zero--mode related to dilatations is 
\beq
\d_{0}A=\frac{\p A}{\p t}=\frac{it}{\sqrt{t^4-a^4}}
\pmatrix{f^{3}(r)u\s_{z} & f^{1}(r)r\s_{-}\cr\cr
f^{1}(r)r\s_{+}& -f^{3}(r)u\s_{z}}\ \ ,
\eeq
and satisfies the usual background gauge condition $D^{\m}\d_{0}A_{\m}=0$. 
The three zero--modes related to global $SU(2)$ color rotations 
are not transverse, but they 
can be made such by adding a local gauge transformation \cite{BFRM}.
The resulting transverse zero--modes are then
\beq
\d_{a}A_{\m}^{b}=(D_{\m}\th_{a})^{b}\ \ .
\eeq
The four bosonic zero--modes are now orthogonal and their norms are
\beqa
||\d_{0}A||^2&=&\frac{8\pi^2 t^4}{t^4-a^4}\ \ , \nonumber\\
||\d_{1}A||^2&=&||\d_{2}A||^2=8\pi^2 t^2\ \ , \\
||\d_{3}A||^2&=&\frac{8\pi^2 (t^4-a^4)}{t^2}\ \ . \nonumber
\eeqa
Here a subtle point arise: collective coordinates are associated to 
unbroken symmetries of the theory. Namely, 
the related transformations leave the chosen vacuum configuration 
\beq\label{allafaccia}
\langle A_{\m}\rangle=0\ \ ,\qquad 
\langle\phi^{a}\rangle=\O^{a}{}_{b}(v \d^{b3})\ \ 
\eeq
unchanged. When the gauge symmetry is spontaneously broken, we may expect 
that the corresponding collective coordinates should not be taken into 
account. This question has been carefully studied in \cite{SHIFCOL}. 
It was pointed out there that we can act on the family of vacuum 
configurations 
(\ref{mizzica}) in two different ways. Rotations acting from the left 
obviously correspond to $SU(2)$ global color transformations. Rotations 
acting from the right are however also possible (they are called ``flavor" 
rotations in \cite{SHIFCOL}). The crucial observation is that the basis 
(\ref{mizzica}) is left invariant when the two rotations are realized 
by the same matrix $R$. In other words, 
\beq
R^{a}{}_{d}\ \phi_{0}^{d(c)}(R^{T})_{c}{}^{b}=\phi_{0}^{a(b)}\ \ .
\eeq
This $SU(2)$ flavor symmetry exists only in 
the space of classical solutions for the 
Higgs field, but not at the Lagrangian level. In particular 
it acts trivially on the gauge sector and it does not affect 
the structure of the gauge zero--modes. 
In flat space the previous observations have no effect; in our case, 
however, we are led to consider the solution of the scalar Laplace 
equation (\ref{scalar}) with boundary conditions dictated 
by (\ref{mizzica2}). A basis of three independent solutions is given by 
\beq
\phi^{a(b)}(x)=\phi_{0}^{a(b)}\th^{a}(x)=v\Big[ \th^{1}\d^{ab}+
(\th^{3}-\th^{1})\d^{a3}\d^{b3}\Big]\ \ . 
\eeq
The point is that, in order to ensure a correct vacuum alignment, 
one is forced to 
act on $\phi^{a(b)}$ 
with the rotations  mentioned above, that is 
\beq
\phi^{a(b)}\longrightarrow 
R^{a}{}_{d}\ \phi^{d(c)}(R^{T})_{c}{}^{b}\equiv
\phi_{R}^{a(b)}\ \ .
\eeq
If we choose the boundary condition 
\beq 
\label{viviana}
\langle\phi^{a}\rangle =  v \d^{a3}
\eeq
for the Higgs field, the correct solution for finite $x$ is
\footnote{Different choices of the matrix $\O$ in (\ref{allafaccia}) 
give physically equivalent theories; so we put $\O=\uno$.}
\beq
\label{salute}
\phi^{a}_{\rm cl}(x)=\phi^{a(3)}_{R}(x) \ \ .
\eeq
The non--triviality of (\ref{salute}) resides in the different expressions 
of the $\th^{a}$ which are a consequence of the isometry group  
of the Eguchi--Hanson manifold. 
On the other hand, in flat space one has 
$\th^{a}_{\rm flat}=x^2 /(x^2+\r^2)$, $a=1,2,3$ (in the singular gauge) and 
the corresponding expression (\ref{salute}) 
for $\phi^{a}_{\rm cl}$ does not contain the matrix $R$ anymore.\\
The expression (\ref{salute}) could also be obtained in a more direct way
\footnote{We thank Gian Carlo Rossi for discussions on this point.}.   
The scalar field configuration $\phi^{a}_{\rm cl}(x)$ can actually be found
by simply requiring that it satisfies the scalar Laplace equation 
in the background of the most general 
({\it i.e.} gauge--rotated)
instanton configuration,  
$RA$, that is 
\beq
\label{velia}
[D^2 (R A)]_{a}{}^{b}\phi^{b}_{\rm cl}=0 \ \ ,
\eeq
with the 
boundary condition (\ref{viviana}). 
Since (\ref{velia}) is equivalent to 
\beq
[D^2 (A)]_{a}{}^{b}(R^{T} \phi_{\rm cl})^{b}=0 \ \ ,
\eeq
we can immediately convince ourselves that (\ref{salute}) 
satisfies (\ref{velia}) and (\ref{viviana}). \\
The matrix $R$ can be written in terms of three Euler angles 
$\th,\varphi,\psi$. As explained before, these angles are in fact the 
global color collective coordinates related to $SU(2)/\Z_{2}$. 
This is the way these instanton moduli come into play in this context.

With these elements we can now calculate the contribution of 
the Higgs field  configuration to the classical action, which 
reads
\beq\label{etciu}
S_{\rm cl}= \frac{1}{g^2}\int d^4 x  \sqrt{g} 
\left[(D^{\m}\phi_{\rm cl}^{\dagger})^a 
(D_{\m}\phi_{\rm cl})^a\right]=
\frac{2\pi^2 t^2 |v|^2}{g^2}\left(1-\frac{a^4}{t^4}\cos^{2}\th\right)\ \ .
\eeq
Note in (\ref{etciu}) the explicit dependence on the gauge 
orientations. 
 
Let us now calculate the Yukawa action $S_{\rm Y}$ 
written with the complete expansion 
of the fermionic  fields replaced by their   
projection over the zero--mode 
subspace. 
According to the
index theorem for the Dirac operator in the background of a 
self--dual gauge field configuration, 
we have only zero--modes of one
chirality, so $S_{\rm Y}$ reduces to 
\beq
\label{yuk}
S_{\rm Y} \left[ \phi ,  \phi^{\dagger}  , \lambda^{(0)},
\bar\lambda=0 \right]=\frac{\sqrt{2}}{g^2}\int d^4 x \sqrt{g}\ \eps_{abc}
(\phi^{\dagger}_{\rm cl})^a 
(\l_{(0)}^b \psi_{(0)}^c)\ \ ,
\eeq
where, for the sake of clarity, we adopted different symbols for the two 
gauginos, $\l=\l_1$ and $\psi=\l_2$. 
Inserting (\ref{gaugino}) and (\ref{salute}) in (\ref{yuk}), 
we get 
\beq
S_{Y}=-\frac{2i\sqrt{2}}{g^2}\int d^4 x \sqrt{g}\left[\th_{1}f_{1}f_{3}(\d_{b1}
w_{1}^{\ast}+\d_{b2}w_{2}^{\ast})+\th_{3}(f_{1})^2 \d_{b3}w_{3}^{\ast}
\right](\eta_{[\l]}\s^{b}\eta_{[\psi]})\ \ ,
\eeq
where $\eta_{[\l]}=\eta_1$ and $\eta_{[\psi]}=\eta_2$ and we have defined
\beqa 
w_{1}&=&v\sin\th\sin\psi\ \ ,\nonumber\\
w_{2}&=&v\sin\th\cos\psi\ \ ,\nonumber\\
w_{3}&=&v\cos\th\ \ .
\eeqa
After a straightforward integration we finally obtain
\beq
S_{Y}=\eta_{[\l]}\s^{b}M^{b\ast}\eta_{[\psi]}\equiv
\frac{-i\sqrt{2}\pi^{2}}{g^2}(\eta_{[\l]}\s^{b}\eta_{[\psi]})\left[
(\d_{b1}w_{1}^{\ast}+\d_{b2}w_{2}^{\ast})\sqrt{t^4-a^4}+
\d_{b3}w_{3}^{\ast}\frac{a^4+t^4}{t^2}\right]\ \ .
\eeq
\section{Computation of instanton--dominated Green's \newline functions}
\setcounter{equation}{0}
Let us now compute the simplest non--zero correlator in the Eguchi--Hanson 
background. 
Since the base manifold is not translationally invariant, there are no 
supersymmetric gaugino zero--modes, unlike the case of flat space. 
Moreover, as there is a non--zero 
vacuum expectation value for the scalar field, the superconformal gaugino 
zero--modes are lifted.  
The integration over the fermionic collective coordinates is thus entirely 
saturated by the Yukawa action,  and the simplest non--zero correlator is 
$\langle\uno\rangle$, {\it i.e.} the partition function itself.
The integration over the bosonic zero--modes is then replaced by an integration 
over the moduli of the instanton. The corresponding 
Jacobian is \cite{BFRM2}
\beq
J=\prod_{I=0}^{3}\frac{||\d_{I}A||}{\sqrt{2\pi}}=\frac{64\pi^4 t^3}
{(\sqrt{2\pi})^4}\ \ ,
\eeq
which does not depend on $a$. This is not an unexpected result,  
since the metric on the minimal instanton moduli space coincides with the 
Eguchi--Hanson metric \cite{BFRM}. 

The evaluation of the Green's function 
$\langle\uno\rangle$ in the semiclassical approximation 
yields, after integrating over non--zero mode fluctuations, 
\beqa
\langle\uno\rangle&=&e^{-\frac{8\pi^2}{g^2}\frac{1}{2}}\m^2\int_{a}^{\infty}
dt\int_{SU(2)/Z_{2}}d^3 \S \frac{64\pi^4 t^3}{(\sqrt{2\pi})^4}
\left(\frac{1}{\sqrt{2}\pi t}\right)^4 \det M^{\ast}\times\nonumber\\
&\times&\exp\left[-\frac{2\pi^2 t^2 |v|^2}{g^2}\left(1-\frac{a^4}
{t^4}\cos^2 \th\right)\right]\ \ ,
\eeqa
where 
$d^3 \S=\frac{1}{8}\sin\th d\th d\varphi d\psi$, 
$M^{\ast}=\s^{b}M^{b\ast}$ and $\det M^{\ast}$ comes 
from integrating $\exp(-S_{\rm Y})$.
Furthermore, $\m^{4-\frac{1}{2}(2+2)} e^{- \frac{8 \pi^2}{2g^2}}
= \L^{2}$, where $\L$  is
the $N=2$  SYM renormalization group invariant scale
with gauge group $SU(2)$. The scale $\m$ comes from the Pauli--Villars
regularization of the determinants, 
and the exponent is
$b_1 c_{2}({\cal E})=n_B-n_F/2$, where $n_B,n_F$ are the number of 
bosonic and fermionic zero--modes and $b_{1}$ is the first coefficient of the 
$\beta$--function of the theory. 
Finally, the factor $\exp ( - 8 \pi^2 / 2g^2 )$ comes from 
the instanton action. 
Writing the determinant explicitly we 
obtain:
\beq
\langle\uno\rangle=-\frac{4\pi^4 (v^{\ast})^2\L^{2}}{g^4} I(a)\ \ ,
\eeq
where 
\beq
I(a)=\int_{a}^{\infty}\frac{dt}{t}\int_{-1}^{1}dy \ 
e^{-\frac{2\pi^2 |v|^2 t^2}{g^2}(1-
y^2 \frac{a^4}{t^4})}\left[ 
(1-y^2)(t^4-a^4)+y^2 \left(\frac{a^4+t^4}{t^2}\right)^2\right]
\ \ ,
\eeq
and $y=\cos\th$. 
In the limit $a\to 0$ in which the Eguchi--Hanson manifold approaches 
the orbifold $\R4/\Z_{2}$, the integral becomes
\beq
\label{azero}
I(a\to 0)=\frac{g^4}{4\pi^4 |v|^4}\ \ ,
\eeq
so that  
\beq
\langle \uno\rangle_{a\to 0}=-\frac{\L^2}{v^2}
\ \ .
\eeq
In the general case $a\neq 0$, after some algebraic manipulations, one gets 
\beqa
\label{anonzero}
I(a)&=&\sum_{n=0}^{\infty}\frac{a^4 x^{2n-2}}{n! (2n+1)}\Gamma(2-n,x)+
\sum_{n=0}^{\infty}\frac{4n a^4 x^{2n}}{n! (2n+1)(2n+3)}\Gamma(-n,x)+
\nonumber\\
&+&\sum_{n=0}^{\infty}\frac{a^4 x^{2n+2}}{n! (2n+3)}\Gamma(-2-n,x)\ \ ,
\eeqa
where $x=2\pi^2 a^2 |v|^2 /g^2$, and $\G(n,x)$ is Euler's incomplete gamma 
function. 
In the limit $a\to 0$ we recover (\ref{azero}). 

As one could expect, the correlator explicitly depends on the Eguchi--Hanson 
parameter $a$. 
Indeed, the dependence on $\L$  is completely fixed by zero--modes 
counting to 
be $\L^{n_{B}-\frac{1}{2}n_{F}}$ (in the present case $n_{B}=n_{F}=4$).  
In the case $v\neq 0$, however, one can form the adimensional quantity 
$a|v|$. Therefore, Ward--Takahashi identities and dimensional analysis 
do not completely fix the correlator dependence on $v$ and $a$. On the 
other hand, when $v=0$, the same argument dictates the independence 
of the correlators on the inverse mass scale $a$.\\
We now extend our analysis to the 
Green's function  $\langle {\rm Tr}\phi^2\rangle$. 
The quantum fluctuations of the complex scalar field are replaced, after 
functional integration, by $\phi_{\rm inh}$, where
\beq 
\phi_{\rm inh}^{a}= \sqrt{2}\ \eps^{bdc} 
[( D^{2})^{-1}]^{ab} (\l_{(0)}^{d}  
\psi_{(0)}^{c}) \ \ .
\label{fiino}
\eeq
So, in this case we need one more ingredient, that is 
the solution of the equations of motion for the scalar field in the 
gaugino zero--modes background 
\beq\label{azzarola}
(D^2\phi_{\rm inh})^{a}=\sqrt{2}\ \eps^{a}{}_{bc}
(\l_{(0)}^{b}\psi_{(0)}^{c})\ \ .
\eeq
The solution of (\ref{azzarola}) is given by 
\beq\label{azzarola2}
\phi^{b}_{\rm inh}=-2i\sqrt{2}h^{b}(\eta_{[\l]}\s^{b}\eta_{[\psi]})
\equiv m^b(\eta_{[\l]}\s^{b}\eta_{[\psi]})\ \ ,
\eeq
where the index $b$ is not summed over and the functions $h^{b}$ are 
\beqa
h^{1}&=&h^{2}=-\frac{\sqrt{(t^4-a^4)(r^4-a^4)}}{4(r^2+t^2)^2}\ \ ,\nonumber\\
h^3&=&-\frac{(a^4+t^4)r^2+2a^4 t^2}{4t^2 (r^2+t^2)^2}\ \ .
\eeqa 
Let us now start our calculation. Unlike  the case $v=0$ there is more than 
one contribution from the insertion of the operator ${\rm Tr}\phi^2$ 
in the functional integral. In fact, all the 
remaining superconformal gaugino zero--modes are lifted, so that also the 
partition function gets a contribution from the instanton background. 
If we separate the classical and the quantum contribution from $\phi^b$ we 
obtain that the integration over the fermionic collective coordinates 
becomes 
\beqa\label{trphi2}
& &\int d^2 \eta_{[\l]} d^2 \eta_{[\psi]}\ e^{-S_{\rm Y}}(\phi_{\rm cl}^b+
\phi_{\rm inh}^b)(\phi_{\rm cl}^b+
\phi_{\rm inh}^b)=\nonumber\\
& &=(\det M^{\ast})(\phi_{\rm cl}^b\phi_{\rm cl}^b)
+2 m^b m^b -4 (\phi_{\rm cl}^b m^b M^{b\ast})
\ \ ,
\eeqa
where $m^{b}$ is defined in (\ref{azzarola2}).
Due to the extreme complexity of the integrals, we limited 
ourselves to study the $a\to 0$ limit, in which (\ref{trphi2}) becomes
\beqa
\widehat{I}(r,t)&=&\int d^2 \eta_{[\l]} d^2 \eta_{[\psi]}\ e^{-S_{Y}}
(\phi_{\rm cl}^b+
\phi_{\rm inh}^b)(\phi_{\rm cl}^b+
\phi_{\rm inh}^b)=\nonumber\\
&=&-\frac{r^4}{(r^2+t^2)^2} \left(\frac{2\pi^4 |v|^4}{g^4}+
\frac{4\pi^2 |v|^2 }{g^2 (r^2+t^2)}+\frac{3}{(r^2+t^2)^2}\right)\ \ .
\eeqa
The correlator has thus the form
\beq\label{fi2}
\langle {\rm Tr}\phi^2\rangle_{a\to 0}=\frac{1}{2}\L^{2}\int_{0}^{\infty}
d t \int_{SU(2)/Z_{2}}d^3 \S\ \frac{64\pi^4 t^3}{(\sqrt{2\pi})^4}
\frac{1}{(\sqrt{2}\pi t)^4}e^{-\frac{2\pi^2 t^2 |v|^2}{g^2}} 
\widehat{I}(r,t)=
\frac{\L^2}{2}\ \  , 
\eeq
and is position--independent, as required by supersymmetry. 
Dimensional analysis and zero--modes counting 
completely constrains the correlator in (\ref{fi2}) to be independent 
of the mass scale $v$.
\section{Discussion}
\setcounter{equation}{0}
In this paper we have computed some correlators in an $SU(2)$, 
$N=2$ SYM on ALE 
backgrounds. 
The choice of a non--zero vacuum expectation value for the scalar field 
introduces a new scale $v$. On dimensional grounds,
when $v=0$, instanton--dominated
correlators obviously do not depend on $a$ \cite{BFRM2}.
However, the study of instanton effects in the low--energy theory 
requires setting $v\neq 0$ from the beginning. 
The Green's functions we have studied 
show an explicit dependence  on the adimensional quantity $a|v|$.
Generally speaking, we expect all correlators to depend on this 
quantity. 
This means that an extension of our calculations to supergravity,  
where $a$ becomes itself a collective coordinate to integrate over, 
is not straightforward. Indeed, we cannot appeal to an explicit  
factorization between the gauge and the gravitational sectors and 
thus exploit the results found in \cite{KONMAGN}.
We want to remark that, as in flat space, the same correlation function has 
different values when calculated in different phases of the theory 
(that is, in the presence or in the absence of a vacuum expectation value 
for the scalar field). 
This point deserves further investigations.

In \cite{MARCO}, microscopic Green's functions were directly related to the 
$N=2$ abelian holomorphic prepotential (in the flat space case). 
It would be interesting to check ({\it e.g.} with instanton techniques) 
if this approach  can be extended to our context.  
We also plan to extend 
our analysis to $N=4$ SYM theories and to the case $c_{2}=1$ in 
a future publication.
 
\vskip 1.5cm
\noindent
{\bf \large Acknowledgements}
\\ \noindent
We would like to thank Francesco Fucito and Gian Carlo Rossi for 
many useful discussions.

\end{document}